\documentclass[sigconf, nonacm]{acmart}
\usepackage{tikz}
\usetikzlibrary{trees,positioning}
\usepackage{subcaption} 
\usepackage{caption}       % For captions
\usepackage{mdframed} % for creating framed boxes
\usepackage[font=small]{subcaption}
\usepackage[font=small]{caption}
\usepackage[ruled,vlined,noend]{algorithm2e}
\usepackage[most]{tcolorbox}
\usepackage{enumitem}
\usepackage{color}
\usepackage{multicol}
\usepackage{multirow}
\usepackage{array}
\usepackage{color, colortbl}
\usepackage{booktabs}
\usepackage{rotating}
\usepackage{stmaryrd}
\usepackage{anyfontsize}
\usepackage{amsthm}
\newtheorem{example}{Example}
\usepackage{soul}
\usepackage{soul}
\usepackage{hyperref}
\usepackage{url}
\usepackage{relsize}
\usepackage[normalem]{ulem}
\hypersetup{
  colorlinks,
  urlcolor=blue}

\definecolor{myred}{RGB}{153,0,0} % Red text color
\definecolor{myblue}{RGB}{0,0,153} % Blue text color
\definecolor{lightgray}{RGB}{230,230,230} % Light gray background color

\newtcolorbox{mybox}{
  colback=gray!15, 
  colframe=black,
  arc=5mm, 
  boxrule=0pt, 
  boxsep=5pt, 
  left=5pt, 
  right=5pt, 
  top=5pt, 
  bottom=5pt 
}
\newcommand\vldbdoi{XX.XX/XXX.XX}

\newcommand\vldbavailabilityurl{}
\newcommand\vldbpagestyle{plain} 

\newcommand{\charles}{\textsc{ChARLES}\xspace}
\newcommand{\ChARLES}{\textsc{ChARLES}\xspace}
\newcommand{\SemDataDiff}{\textsc{ChARLES}\xspace}

\newcommand{\datadiff}{\textsc{ChARLES}\xspace}
\newcommand{\stepCounter}[1]{{\Large\textcircled{{\small#1}}}}

\begin{document}
\title[\ChARLES: Change-Aware Recovery of Latent Evolution Semantics in Relational Data]
	{\ChARLES: Change-Aware Recovery of Latent Evolution Semantics in Relational Data}

%%
%% The "author" command and its associated commands are used to define the authors and their affiliations.
\author{Shiyi He}
\affiliation{%
  \institution{University of Utah}
  \streetaddress{P.O. Box 1212}
  \city{Salt Lake City}
  \country{USA}
  \postcode{43017-6221}
}
\email{shiyi.he@utah.edu}

\author{Alexandra Meliou}
\affiliation{%
  \institution{University of Massachusetts Amherst}
  \city{Amherst}
  % \state{Massachusetts}
  \country{USA}}
\email{ameli@cs.umass.edu}

\author{Anna Fariha}
\affiliation{%
  \institution{University of Utah}
  \city{Salt Lake City}
  % \state{Utah}
  \country{USA}}
\email{afariha@cs.utah.edu}

\begin{abstract}
	\looseness-1 Data-driven decision-making is at the core of many modern
applications, and understanding the data is critical in supporting trust in
these decisions. However, data is dynamic and evolving, just like the
real-world entities it represents. Thus, an important component of
understanding data is analyzing and drawing insights from the changes it
undergoes. Existing methods for exploring data change list differences
exhaustively, which are not interpretable by humans and lack salient insights
regarding change trends. For example, an explanation that \emph{semantically}
summarizes changes to highlight gender disparities in performance rewards is
more human-consumable than a long list of employee salary changes.

% \looseness-1 Motivated by these challenges, we present \ChARLES, a system for
% explaining \emph{semantic} differences between two snapshots of an evolving
% database, in an effective, concise, and interpretable way. Our key observation
% is that changes in databases are typically systematic, resulting from a series
% of batch updates over different data partitions. We leverage this observation
% to generate a ``change explanation''---a set of transformations that can be
% applied to an earlier version of the database to obtain the later
% version---that ensures that (1)~the explanation accurately captures the most
% salient changes in the data and (2)~humans can reasonably interpret the
% explanation. Under the hood, \ChARLES compares database versions, infers
% feasible transformations by fitting multiple regression lines over different
% partitions of the data, and introduces a ranking function to rank the
% explanations balancing the trade-offs between accuracy and interpretability.
% Through an intuitive and interactive interface, \ChARLES enables the user to
% customize the system to obtain their preferred explanation. We demonstrate how
% \ChARLES can help users effectively reason about data evolution over
% real-world datasets.

\looseness-1 We demonstrate \ChARLES, a system that derives \emph{semantic}
summaries of changes between two snapshots of an evolving database, in an
effective, concise, and interpretable way. Our key observation is that, while
datasets often evolve through point and other small-batch updates, rich data
features can reveal \emph{latent semantics} that can intuitively summarize the
changes. Under the hood, \ChARLES compares database versions, infers feasible
transformations by fitting multiple regression lines over different data
partitions to derive change summaries, and ranks them. \charles allows users to
customize it to obtain their preferred explanation by navigating the
accuracy-interpretability tradeoff, and offers a proof of concept for reasoning
about data evolution over real-world datasets.

\end{abstract}

\maketitle

\pagestyle{\vldbpagestyle}

% \begingroup\small\noindent\raggedright\textbf{PVLDB Reference Format:}\\
% \vldbauthors. \vldbtitle. PVLDB, \vldbvolume(\vldbissue): \vldbpages, \vldbyear.\\
% \href{https://doi.org/\vldbdoi}{doi:\vldbdoi}
% \endgroup
% \begingroup
% \renewcommand\thefootnote{}\footnote{\noindent
% This work is licensed under the Creative Commons BY-NC-ND 4.0 International License. Visit \url{https://creativecommons.org/licenses/by-nc-nd/4.0/} to view a copy of this license. For any use beyond those covered by this license, obtain permission by emailing \href{mailto:info@vldb.org}{info@vldb.org}. Copyright is held by the owner/author(s). Publication rights licensed to the VLDB Endowment. \\
% \raggedright Proceedings of the VLDB Endowment, Vol. \vldbvolume, No. \vldbissue\ %
% ISSN 2150-8097. \\
% \href{https://doi.org/\vldbdoi}{doi:\vldbdoi} \\
% }\addtocounter{footnote}{-1}\endgroup
% %%% VLDB block end %%%

\ifdefempty{\vldbavailabilityurl}{}{
\vspace{.3cm}
\begingroup\small\noindent\raggedright\textbf{PVLDB Artifact Availability:}\\
The source code, data, and/or other artifacts have been made available at \url{\vldbavailabilityurl}.
\endgroup
}

\section{Introduction}

\begin{figure}[t]
\centering

\begin{subfigure}{0.485\linewidth}
\centering
\resizebox{1\textwidth}{!}{
\begin{tabular}[t]{@{}lll@{ }c@{ }rr@{}}
\toprule
\textbf{name} & \textbf{gen} & \textbf{edu} & \textbf{exp} & \textbf{salary} & \textbf{bonus} \\
\midrule
Anne 	& F & PhD 	& 2 & \$230,000 & \$23,000 \\
Bob 	& M & PhD 	& 3 & \$250,000 & \$25,000 \\
Amber 	& F & MS 	& 5 & \$160,000 & \$16,000 \\
Allen 	& M & MS 	& 1 & \$130,000 & \$13,000 \\
Cathy 	& F & BS 	& 2 & \$110,000 & \$11,000 \\
Tom 	& M & MS 	& 4 & \$150,000 & \$15,000 \\
James 	& M & BS 	& 3 & \$120,000 & \$12,000 \\
Lucy 	& F & MS 	& 4 & \$150,000 & \$15,000 \\
Frank 	& M & PhD 	& 1 & \$210,000 & \$21,000 \\
\bottomrule
\end{tabular}
}
\vspace{-2mm}
\caption{2016 snapshot}
\label{fig:firsttable}
\end{subfigure}
\hfill
\begin{subfigure}{0.485\linewidth}
\centering
\resizebox{1\textwidth}{!}{
\begin{tabular}[t]{@{}lll@{ }c@{ }rr@{}}
\toprule
\textbf{name} & \textbf{gen} & \textbf{edu} & \textbf{exp} & \textbf{salary} & \textbf{bonus} \\
\midrule 
Anne 	& F & PhD	& 3 	& \$230,000 & \hl{\$25,150} \\
Bob 	& M & PhD	& 4 	& \$250,000 & \hl{\$27,250} \\
Amber 	& F & MS 	& 6 	& \$160,000 & \hl{\$17,440} \\
Allen 	& M & MS 	& 2 	& \$130,000 & \hl{\$13,790} \\
Cathy 	& F & BS 	& 3 	& \$110,000 & \$11,000 \\
Tom 	& M & MS 	& 5 	& \$150,000 & \hl{\$16,400} \\
James 	& M & BS 	& 4 	& \$120,000 & \$12,000 \\
Lucy 	& F & MS 	& 5 	& \$150,000 & \hl{\$16,400} \\
Frank 	& M & PhD	& 2 	& \$210,000 & \hl{\$23,050} \\
\bottomrule
\end{tabular}
}
\vspace{-2mm}
\caption{2017 snapshot}
\label{fig:secondtable}
\end{subfigure}
\vspace{-3mm} 
\caption{\small{Employee salaries have evolved over a year, with the
\emph{bonus} attribute increasing by 8--10\% (highlighted in yellow). Context
and trends of these changes are not apparent from the point updates.}}
\vspace{-5mm}
\label{fig:table}
\end{figure}

The task of data understanding is of prime importance in today’s data-driven
world. To make sense of data, existing data summarization systems enable users
to interactively understand the content of a static
database~\cite{DBLP:journals/tkde/JoglekarGP19}. However, data is dynamic,
evolving over time, and summarization techniques for static databases are
ineffective at explaining this data evolution. Datasets often evolve through
many tuple-level or small-batch-level updates. However, exhaustively listing
all such fine-grained changes overwhelms human analysts. Fortunately, rich
features in the data have the potential to concisely summarize such
fine-grained changes, offering an ``explanation'' that captures the salient
\emph{semantics} of the data's evolution.

\looseness-1 To understand change, we need to understand its mechanism,
quantification, and cause. This information is hard to extract from change
logs, which are often unavailable or inaccessible to end users. Even if
available, their format is often unsuitable for consumption by non-experts,
and, thus, interpreting such large change logs to understand data changes poses
a significant hurdle for data consumers. Data versioning techniques can trace
the locations and quantities of changes, but high-level trends are not
typically obvious at that fine granularity. Instead, changes should be
summarized at a coarser granularity to reveal the underlying causes and
mechanisms.

\begin{example}\label{ex:one}
	 \looseness-1 Figure~\ref{fig:table} presents two snapshots of a salary
	 database in (a)~2016 and~(b)~2017. In 2016, \texttt{bonus} was a flat 10\% of
	 \texttt{salary} for all employees. In contrast, we observed no such
	 straightforward trend in 2017. In some cases, the value of \texttt{bonus}
	 differs from last year's value (highlighted in yellow), while in some cases
	 they are identical (for Cathy and James). Furthermore, the difference ranges
	 from 8\% to 10\% and is not identical for everyone. Simply knowing that
	 \texttt{bonus} changed from last year leaves one unsatisfied, as it is not
	 obvious what is the underlying trend behind such non-uniform changes.
	 
	 It turns out that the company opted for a policy to reward long-serving
	 employees and promote educational advancement. In 2017, the company decided
	 to depart from a flat-rate bonus to a customized one. The new scheme for
	 bonus calculation is influenced by three principles: (1)~no one should
	 receive lower bonus than the previous year, (2)~employees with higher level
	 of education should be rewarded more, and (3)~long-serving employees should
	 be rewarded more.
	 
	 This is not immediately apparent by just looking at the data, since
	 \texttt{bonus} for 2017 is no longer directly tied to \texttt{salary}, as was
	 the case in 2016. Instead, it is calculated based on a combination of last
	 year's \texttt{bonus}, employee's education (\texttt{edu}), and years of
	 experience (\texttt{exp}). Specifically, the following rules accurately
	 explain the change trend:

    \begin{itemize}
    
		\item \textbf{R1}: Employees who have a PhD receive a 5\% increase on
		last year's bonus, plus flat \$1000.
    
		\item \textbf{R2}: Employees who have an MS and served for at least 3
		years receive a 4\% increase on last year's bonus, plus flat \$800.

		\item \textbf{R3}: Employees who have an MS and served for less than 3
		years receive a 3\% increase on last year's bonus, plus flat \$400.
			
    \end{itemize}
\end{example}

There are two desirable properties for a \emph{change summary}: (1)~it should
be \emph{precise}, i.e., be able to explain the changes \emph{accurately}, and
(2)~it should be \emph{interpretable} and \emph{succinct} for easy human
consumption. Note that there is a natural tension between these two desirable
properties. Consider the following change summary:
\begin{itemize}
	\item \emph{\textbf{R4}: Everyone receives about 6\% increase on last
year's bonus.}
\end{itemize}
\textbf{R4} is more interpretable (more succinct and human-consumable) than
$\{$\textbf{R1}, \textbf{R2}, \textbf{R3}$\}$; however, \textbf{R4} does not accurately 
capture the change, while $\{$\textbf{R1}, \textbf{R2}, \textbf{R3}$\}$ does. In contrast, 
one can provide a change summary by listing each individual cell that changed. However, 
such a summary---despite being very precise---would lack interpretability as this level 
of detail overwhelms the user.

\looseness-1 \subsubsection*{\datadiff} To meet the requirements of accuracy
and interpretability, we developed \ChARLES
(\underline{Ch}ange-\underline{A}ware \underline{R}ecovery of
\underline{L}atent \underline{E}volution \underline{S}emantics), a system for
producing a \emph{semantic summary of changes} between two snapshots of a
relational database, while striking a balance between accuracy and
interpretability. Our key observation is that data changes are often driven by
some underlying policies and the patterns within data evolution, as manifested
by the changes, can potentially recover those policies. In this work, we focus
on temporal changes and assume that, given a \emph{source} dataset (earlier
version) and a \emph{target} dataset (later version) of identical schema, the
latter is obtained via a set of update operations over the former. Furthermore,
we assume that no tuples are inserted or deleted; only (numerical) values of
various cells are altered.

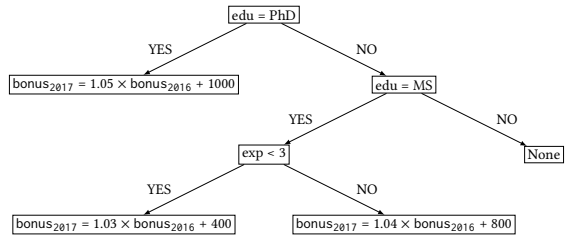
\begin{figure}[t]
\centering
\resizebox{0.42\textwidth}{!}{
\begin{tikzpicture}[
    every node/.style={align=center},
    level 1/.style={sibling distance=60mm},
    level 2/.style={sibling distance=60mm, level distance=15mm},
    level 3/.style={sibling distance=60mm, level distance=15mm}, 
    edge from parent/.style={draw, -latex}
]
  \node[draw,rectangle,inner sep=2pt] {edu = PhD}
    child {node[draw,rectangle,inner sep=2pt] {$\mathtt{bonus_{2017}}$ = 1.05 $\times$ $\mathtt{bonus_{2016}}$ + 1000}
        edge from parent node[left, pos=0.5] {YES\phantom{sss}}}
    child {node[draw,rectangle,inner sep=2pt] {edu = MS}
      child {node[draw,rectangle,inner sep=2pt] {exp < 3}
        child {node[draw,rectangle,inner sep=2pt] {$\mathtt{bonus_{2017}}$ = 1.03 $\times$ $\mathtt{bonus_{2016}}$ + 400}
          edge from parent node[left, pos=0.5] {YES\phantom{sss}}}
        child {node[draw,rectangle,inner sep=2pt] {$\mathtt{bonus_{2017}}$ = 1.04 $\times$ $\mathtt{bonus_{2016}}$ + 800}
          edge from parent node[right, pos=0.5] {\phantom{sss}NO}}
        edge from parent node[left, pos=0.5] {YES\phantom{sss}}
      }
      child {node[draw,rectangle,inner sep=2pt] {None}
        edge from parent node[right, pos=0.5] {\phantom{sss}NO}}
    edge from parent node[right, pos=0.5] {\phantom{sss}NO}
    };
\end{tikzpicture}
}

\vspace{-3mm}
\caption{{\small A linear model tree explaining diff in datasets in Figure~\ref{fig:table}}.}
\vspace{-5mm}
\label{fig:tree}
\end{figure}

The core challenge in this problem is to derive a partitioning of the tuples,
such that tuples within each partition conform to a uniform ``transformation''
of reasonable complexity. As proof of concept, \ChARLES uses k-means clustering
to guide the search for data partitions based on a subset of data attributes
(e.g., education and year of experience). Once approximate partitions are
discovered, \ChARLES applies linear regression to find the most suitable
transformations to capture the changes within each partition (e.g.,
$\mathtt{bonus_{2017}} = 1.05 \times \mathtt{bonus_{2016}} + 1000$). The output
is a \emph{linear model tree}~\cite{DBLP:conf/icml/Potts04}
(Figure~\ref{fig:tree}), where the path from the root to a leaf defines a
partition and the leaf defines the transformation (a linear model).

Furthermore, \ChARLES enhances user experience by
(1)~\emph{customization}---users can specify system parameters such as the
maximum number of attributes they want to see in the change summary---and
(2)~\emph{visualization}---they can interactively inspect different partitions
of the data and the corresponding change trends.

\smallskip

\looseness-1 \noindent\emph{Limitations.} \ChARLES focuses on finding an
interpretable summary of data changes based only on the data, without any
knowledge of external information. While the change summary produced by
\charles may not always match the factual explanation (e.g., when change is due
to some external factors), it nevertheless helps facilitate the development of
hypotheses about the underlying causes of these changes. While \charles relies
on linear models to capture change trends, this can be extended by augmenting
the data with nonlinear features. However, nonlinear models are less
interpretable, which justifies our choice of linear models.

\smallskip

\noindent\emph{Related work.}
Prior work~\cite{DBLP:journals/pvldb/BleifussBJKNS18} has studied the problem
of exploring the entire history of changes in a database, but is limited to
syntactic or raw changes, suitable for historical change exploration involving
a particular entity. Database comparator tools, such as
PostgresCompare~\cite{postgrescompare}
% \footnote{PostgresCompare: \url{www.postgrescompare.com/}} 
and
RDBMS version control system OrpheusDB~\cite{DBLP:journals/pvldb/HuangXLEP17},
% \footnote{OrpheusDB: \url{https://orpheus-db.github.io/}}, 
only look for syntactic changes---values are changed, objects are
altered, rows are removed or added---which is not concise enough to provide 
high-level insights of the changes. Data-diff~\cite{DBLP:conf/kdd/SuttonHGC18} explores
change in distributions of datasets, specialized in the context of data
wrangling. Muller et al.~\cite{DBLP:conf/cikm/MullerFL06} describe change in
two datasets in terms of ``update distance'', defined by the minimal number of
insert, delete, and modification operations necessary. However, none of these
works focus on summarizing changes between two databases.

Explain-Da-V~\cite{DBLP:journals/pvldb/ShragaM23} is closest to \ChARLES as it
also explains transformations that convert a source dataset to a target
dataset. However, it focuses on semantics of schema and data format
transformations (e.g., data extraction, row deletion, adding attributes
representing length of an attribute). In contrast, we focus on semantic changes
where values of attributes are altered based on interactions among other
attributes, with the goal of explaining the evolution of real world entities
represented by datasets.

Local explanations for ML models~\cite{DBLP:conf/kdd/Ribeiro0G16} is
inapplicable in our use case as they focus on classification. In contrast, we
focus on the challenging problem of \emph{pattern-based clustering}, where we
need to find clusters where elements share the same change patterns. However, a
natural cyclic dependency exists in finding shared change patterns and
clustering, as shared patterns can only be discovered once clusters are formed,
where the clusters must be formed such that elements within the same cluster
exhibit identical change patterns.

In our demonstration, participants will witness how \ChARLES generates change
summaries from two datasets, tailored to user preferences, and enables them to
effectively gain insights about data changes. We proceed to describe our
solution sketch in Section~\ref{sol} and then provide the demonstration outline
in Section~\ref{demo}.

\section{Solution Sketch}\label{sol}

Given two datasets of identical schema---a source dataset \(\mathcal{D}_s\) and
a target dataset \(\mathcal{D}_t\)---and a numerical attribute of interest
$a_i$, we aim to produce a \emph{ranked list} of \emph{change summaries} that
capture the changes observed between \(\mathcal{D}_s(a_i)\) and
\(\mathcal{D}_t(a_i)\). Each change summary consists of a set of
\emph{transformations} over different data partitions. We rank the summaries
based on their \emph{scores}, which indicate how well they can strike the
balance between accuracy and interpretability. We assume that $\mathcal{D}_s$
and $\mathcal{D}_t$ contain the same real-world entities, i.e., only values of
non-primary-key attributes were modified, and there were no insertions or
deletions of tuples.

\noindent \subsubsection*{Change summary and conditional transformation} Our
unit of explanation within a change summary is a \emph{conditional
transformation} (CT), which comprises a \emph{condition} and a
\emph{transformation}. A summary $S = \{CT_1, CT_2, \dots\}$ comprises a set of
CTs. The condition explains why a change happened, and the transformation
describes the change itself. For instance, the following CT explains that
employees with a PhD got 5\% increase in bonus plus $\$1000$.
\begin{equation*} 
	\footnotesize{ \textstyle \underbrace{\text{edu = PhD}}_{\text{Condition}} 
	\rightarrow \underbrace{\text{new\_bonus = 1.05 }\times\text{ old\_bonus + 1000}}_{\text{Transformation}} } 
\end{equation*}

\subsubsection*{Desiderata for change summary} A desirable change summary must
ensure that (1)~all or most of the data changes are sufficiently covered and
(2)~the summary itself is interpretable and succinct for human consumption. To
this end, we introduce $Score(S) \in [0, 1]$ for a summary $S$, which indicates
how well $S$ can represent the differences between \(\mathcal{D}_s(a_i)\) and
\(\mathcal{D}_t(a_i)\). 
\setlength{\abovedisplayskip}{3pt}
\setlength{\belowdisplayskip}{3pt} 
{\begin{equation*} 
\textstyle Score(S) = \alpha \times Accuracy(S) + (1-\alpha)\times Interpretability(S)
\end{equation*} }
%
% \noindent 
Here, we model \emph{Accuracy} by the inverse $L_1$ distance between
$\hat{\mathcal{D}_s}(a_i)$ and $\mathcal{D}_t(a_i)$, where
$\hat{\mathcal{D}_s}$ is the transformed dataset obtained by applying the CTs
in $S$ on $\mathcal{D}_s$, and $\alpha$ is a system parameter that controls the
interplay between accuracy and interpretability.

To achieve interpretability, we prioritize the following characteristics of summaries:
% We use several intuitions to model \emph{Interpretability}:

\begin{itemize}[leftmargin=*]
	 \item \emph{Smaller summaries}. A summary with fewer CTs is
	 preferable, as it leads to increased conciseness.
	
	 \item \emph{Simpler conditions and transformations}. A condition consists
	 of a series of descriptors that identify specific segments of the data, so we
	 prefer a simpler condition with fewer descriptors. E.g., the transformation
	 ``All Female employees received 5\% bonus'' is more interpretable than ``All
	 Asian, European Females, or Females working in HR received a 5\% bonus''.
	 Similarly, a transformation involves a linear equation, so a transformation
	 with fewer variables in the equation is preferred.

	 \item \emph{Conditions with higher data coverage}. A condition that
	 yields a small data partition explains little of the change. Thus,
	 we prefer conditions with higher coverage, yielding larger
	 partitions.

	 \item \emph{Higher ``normality'' for conditions and transformations}.
	 Conditions and transformations may involve numeric constants. We prefer the
	 ones involving more ``normal'' values. E.g., the condition ``Age > 25'' is
	 more normal than ``Age > 23.796'', and $5\%$ for a salary increase is more
	 normal (and interpretable) than $2.479\%$. We rely on domain expertise to
	 learn such notions of normality.
 
\end{itemize}

%  \am{The descriptor of the second bullet is vague (what is ``description'' or ``equation'', and the distinction from the 4th criterion is not clear.}

Enhancing interpretability without significantly compromising accuracy results
in a more effective summary, with a higher overall score. The parameter
$\alpha$ controls the tradeoff between accuracy and interpretability and has a
default value of $0.5$. \charles is designed to cater to a diverse audience,
allowing novices to bypass system parameter tuning and experts to adjust
parameters.

\begin{figure}[t]
% https://www.figma.com/file/ryM9tjl7Jmgci0DiK4h46d/workflow?type=design&node-id=0%3A1&mode=design&t=79Q5mlE9XWBjwoWe-1
	\centering
	\includegraphics[width=0.85\linewidth]{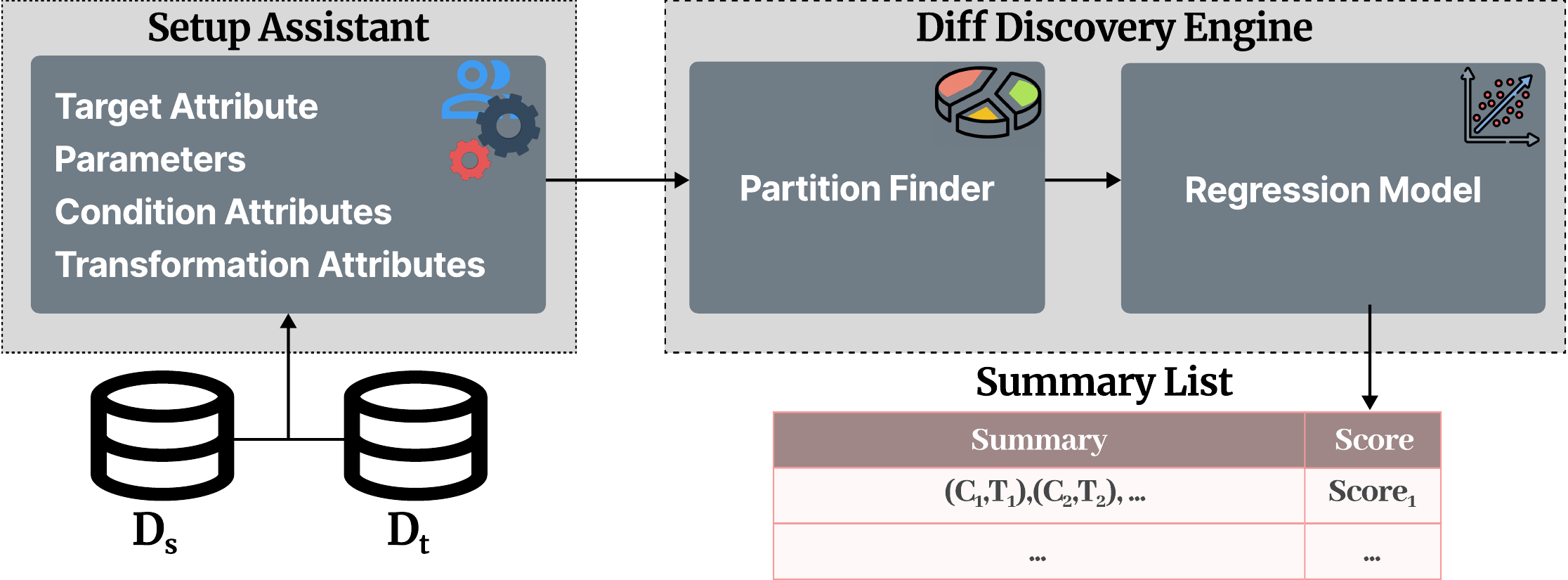}
	\vspace{-2mm}

	\caption{{\small \looseness-1 \SemDataDiff overview: The setup assistant
helps users choose system parameters such as attributes to consider for
conditions and transformations, and the diff discovery engine summarizes
the changes based on data partitioning and fitting regression lines.}}
\vspace{-5mm}
\label{workflow}
\end{figure}

\subsubsection*{\ChARLES architecture} Figure~\ref{workflow} shows the \ChARLES
architecture. It consists of two components: the \emph{setup assistant}, which
helps the users (optionally) tune system parameters, and the \emph{diff
discovery engine}, which is responsible for generating the change summaries.

\smallskip

\noindent \looseness-1 \textbf{Setup assistant.} For datasets with many
attributes, the search space for possible summaries can explode. A user may
help narrow down the set of relevant attributes, which has the dual benefits of
reducing the search space, and pivoting the change summaries around attributes
of interest. However, users may struggle to provide such input when they are
unfamiliar with the schema. \ChARLES addresses this challenge by estimating the
influence of other attributes on the target attribute using correlation
analysis and presents to the user a shortlist of attributes that are most
likely to be effective for explaining the changes. By default, \ChARLES
presents candidate attributes for both condition ($\mathcal{A}_{cond}$) and
transformation ($\mathcal{A}_{tran}$) that have a correlation with the target
attribute greater than 0.5. Additionally, \charles allows the user to narrow
down or expand the candidate attributes by providing two parameters, $c$ and
$t$, where $c$ defines the maximum number of attributes $\in
\mathcal{A}_{cond}$ to use for partition discovery, and $t$ denotes the maximum
number of numerical attributes $\in \mathcal{A}_{tran}$ to use to fit the
linear model within each partition.
 
% \af{You mentioned before that for $\alpha$ we provide default values so novices can bypass parameter tuning, but what about these new parameters $c$ and $t$? How do \charles get these parameters? Who tunes them? When? Are there any default values for these as well?}
% \sh {reply to Anna's comment: We don't have default values for these parameters here, since we will ask the user to decide the complexity of the summary they expect.}\af{But in the demo scenario, you mention 3 as default values.}

\smallskip

\noindent \textbf{Diff discovery engine.} The diff discovery
engine comprises (1)~partition discovery, which identifies potentially
significant partitions, and (2)~transformation discovery, which fits a linear
regression model within each partition. The goal to find a list of change
summaries $SL = \{(S_1,score_1), (S_2,score_2), \ldots\}$, ordered by
decreasing order of their scores.
% ($score_1 \ge score_2 \ge \ldots$).
For each
summary $S_i$, \ChARLES discovers different data partitions specified by
conditions (defined by a subset of attributes in $\mathcal{A}_{cond}$), where
each partition conforms to a specific transformation (defined by a subset of
attributes in $\mathcal{A}_{tran}$).

\begin{figure*}[t]
% https://www.figma.com/file/ytM1oQLYgQlRhrEVytIGoz/Untitled?type=design&node-id=0%3A1&mode=design&t=VHGeVDbZkg5w7mCb-1
	\centering
	\vspace{-4mm}
	\includegraphics[width=0.8\textwidth]{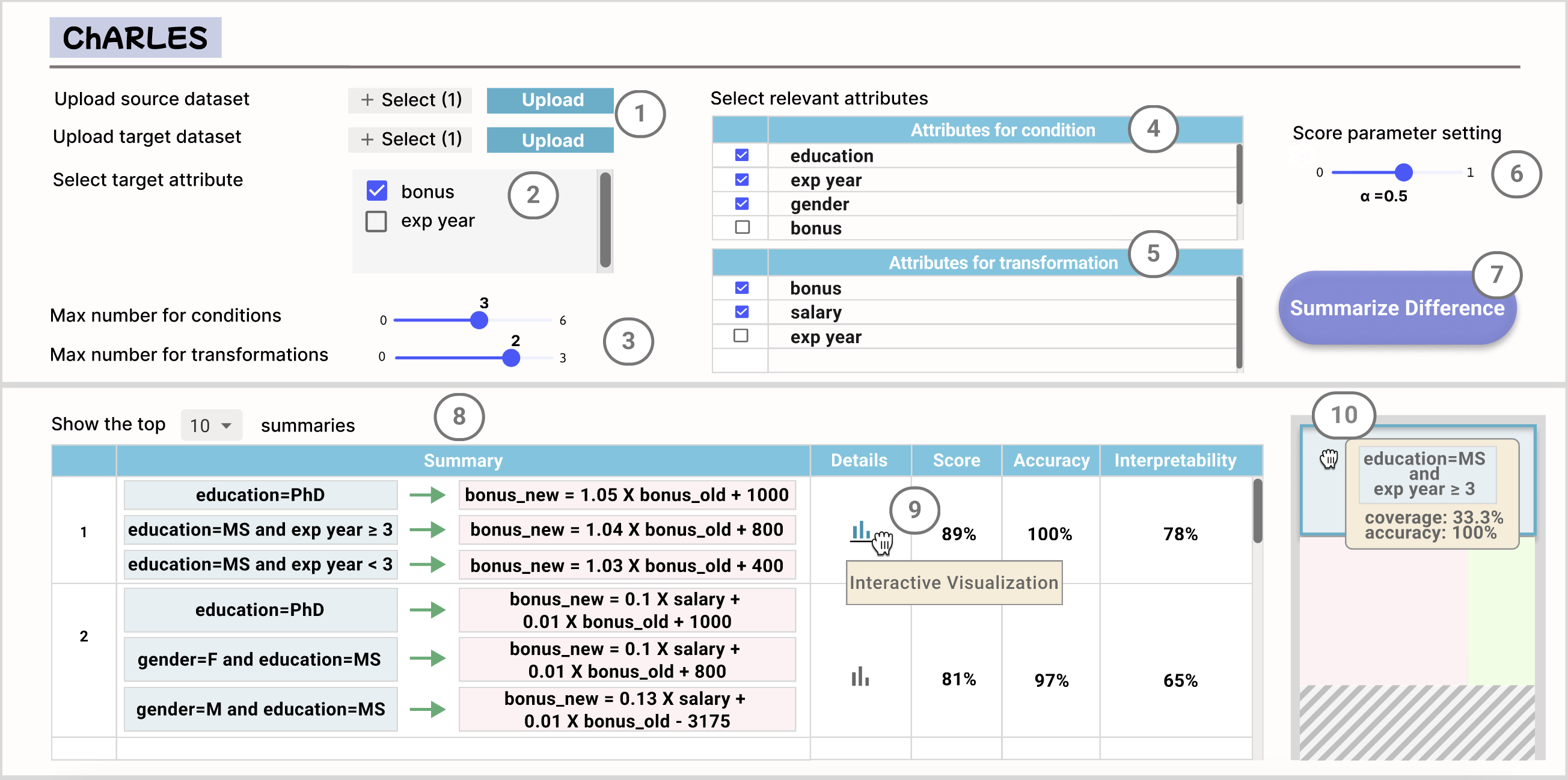}
	 \caption{{\small The \SemDataDiff demo: 
	 \stepCounter{1} upload datasets,
	 \stepCounter{2} select the target attribute,
	 \stepCounter{3} specify the maximum number of attributes for condition and transformation,
	 \stepCounter{4} \ChARLES selects attributes for condition automatically,
	 \stepCounter{5} \ChARLES selects attributes for transformation automatically,
	 \stepCounter{6} tune score parameter $\alpha$,
	 \stepCounter{7} request change summaries,
	 \stepCounter{8}~\ChARLES presents a list of ranked summaries, with their overall scores, and scores for accuracy and interpretability,
	 \stepCounter{9} click on a summary for more details, 
	 \stepCounter{10} detailed visualization of data partitions.}}
	 \vspace{-2mm}
	\label{ui}
\end{figure*}

% \af{Don't mix up tense. We usually use present indefinite
% tense.} 

Based on $\mathcal{A}_{cond}$, $\mathcal{A}_{tran}$, and the
parameters $c$ and $t$ 
% \af{You mentioned $k$ and $c$ above, but talk about $c$
% and $t$. Make a careful pass over your notations.}
, \ChARLES enumerates all
possible combinations of attributes to use for partitioning and generating
transformations. For example, for the parameters $c = 3$ and $t = 2$, \ChARLES
will consider all subsets of $\mathcal{A}_{cond}$ with cardinality $ \le 3$ as
partitioning attributes and all subsets of $\mathcal{A}_{tran}$ with
cardinality $\le 2$ as transformation attributes.
% \af{what are ``regression attributes''? Be
% consistent about names. Previously you only mentioned ``transformation
% attributes''.} 

% \af{You should highlight first that the diff discovery engine comprises two
% parts, partition discovery and transformation discovery. You should also
% mention at a high level how they interact with each other before describing
% these components.}

\smallskip

\looseness-1 \noindent \emph{Partition discovery.} For a specific set of
condition attributes $\mathcal{C} \subseteq \mathcal{A}_{cond}$ and
transformation attributes $\mathcal{T} \subseteq \mathcal{A}_{tran}$, \ChARLES
first fits a linear regression model for $a_i$, over the entire data, based on
the attributes in $\mathcal{T}$. Then \ChARLES performs K-means clustering,
based on the distance from the regression line, to discover potentially
meaningful partitions in terms of the attributes in $\mathcal{C}$.

\smallskip

\looseness-1 \noindent \emph{Transformation discovery.} For each discovered
partition, \ChARLES again fits a linear regression model based on $\mathcal{T}$
to generate a transformation. All such transformations over different
partitions, together, result in a change summary, which can be represented
using a linear model tree similar to Figure~\ref{fig:tree}. Once all summaries
$\{S_1, S_2, \ldots\}$ are generated within the specified parameters, \ChARLES
computes the $Score$ of each summary, ranks the generated summaries according
to the descending order of their scores, and returns the ranked list of
summaries $\{(S_1,score_1), (S_2,score_2), \ldots\}$.

% \af{Fix notations here. we changed it to summary ($S$) and $score$}.

% \paragraph{Transformation discovery.} Clustering provides a general
% partitioning result \af{Again, are you doing clustering or doing Linear tree?
% Both are different algorithms. Pick one and describe it precisely. I recommend
% do the one you did during SIGMOD demo submission and stick to it. No change was
% required in this part anyway. We haven't conducted experiments yet so we don't
% even know if linear tree is viable. So I would recommend just revert back old
% text here.}. Our next step involves performing linear regression on each
% partition again to obtain a linear regression model that better fits each
% specific partition. Once partitions are discovered, for each partition,
% \ChARLES again fits a regression model based on the transformation attributes
% in $\mathcal{T}$ to generate a transformation. All such transformations over
% different partitions, together, result in a change summary, which can be
% modeled using a regression tree similar to the one shown in
% Figure~\ref{fig:tree}. Once all summaries $\{S_1, S_2, \ldots\}$ are generated
% within the specified parameters, \ChARLES computes $Score$ of each summary,
% ranks the generated summaries according to the descending order of their
% respective scores, and returns the ranked list of summaries $\{(S_1,s_1),
% (S_2,s_2), \ldots\}$.

\smallskip

This is a proof-of-concept prototype implementation of \ChARLES. Other methods
of partitioning and transformation discovery are certainly possible, but we
defer deeper investigation to future work.

\section{Demonstration} \label{demo}

We will demonstrate \ChARLES on a real-world dataset~\cite{employeeSalaries}
representing salary information for all active, permanent employees of
Montgomery County, MD for the years 2016 and 2017. The dataset contains
information about employee salaries over $8$ attributes, including Department,
Department Name, Division, Gender, Base Salary, Overtime Pay, Longevity Pay,
and Grade. Figure~\ref{ui} shows \ChARLES's graphical user interface. During
the demonstration, we will guide the participants through ten steps. We have
annotated each step with a circle.

\smallskip \textbf{Step \stepCounter{1} (Uploading datasets)} The user uploads
two dataset versions they want to compare. For ease of exposition, we use the
toy datasets of Example~\ref{ex:one} in this demo scenario.

\textbf{Step \stepCounter{2} (Selecting the target attribute)} Next, the user
chooses the target attribute that manifests changes they wish to investigate.
For our scenario, the user chooses ``bonus''.

\textbf{Step \stepCounter{3} (Setting parameters)} Next, the user chooses the
maximum number of condition attributes to use for partitioning (3) and the
maximum number of transformation attributes (2).

\textbf{Steps \stepCounter{4} \& \stepCounter{5} (Attribute selection)}
\datadiff presents a ranked list of attributes that are most promising for
condition \stepCounter{4} and transformation \stepCounter{5} attributes. Based
on the user specifications, \ChARLES selects the top 3 results from the
condition attributes list and the top 2 results from the transformation
attributes list. Users can accept this default selection or interactively
filter out undesired attributes and select other attributes. In our case, the
user accepts the default selection: ``education'', ``exp year'', and ``gender''
as potential condition attributes and ``bonus'' (of the previous year) and
``salary'' as potential transformation attributes.

% \textbf{Step \stepCounter{6} (Advanced settings)} 
% If the user wishes,
% they can configure advanced system parameters. For example, the default value
% of $\alpha$ is set to $0.5$. However, users have the discretion to either
% accept this default setting or adjust the parameters to meet their specific
% requirements. For example, if they seek a more interpretable explanation, they
% can tune $\alpha$ to a lower value, thereby, prioritizing interpretability over
% accuracy.

% \af{If advanced settings allow to tune only ONE parameter $\alpha$, I'd rather keep it in the main UI and not have yet another button for this. You can simply keep a slider for $\alpha$ in the main UI if that is the ONLY thing one can do in the advanced mode.}
% \af{This is a demo scenario, not a detailed manual of our tool.
% Modify your writing to highlight a scenario. If you want to talk about system
% properties (what we support), probably move it in Section 2. Also, this is too
% verbose, compress it.}

% \looseness-1
\textbf{Steps \stepCounter{6}--\stepCounter{8} (Summarize changes)} The value of
$\alpha$ represents the weight of accuracy in the $Score$ function, which is
set to $0.5$ by default. Users can modify this parameter according to their
requirements \stepCounter{6}. For those aiming for a more interpretable
summary, adjusting $\alpha$ to a lower value can shift the balance towards
interpretability at the expense of accuracy. The user then requests to generate
change summaries \stepCounter{7} and \ChARLES displays the summaries
\stepCounter{8}. Each summary comprises a set of conditional
transformations---where conditions are in light blue and transformations are in
light pink---followed by an option to visualize \stepCounter{9}, overall score,
accuracy, and interpretability scores. In this scenario, the first summary
produced by \ChARLES reflects the scenario described in Example~\ref{ex:one},
which incurs a very high score of 89\%. By default, \ChARLES presents the $10$
top-scoring summaries.
% If fewer than 10 summaries are available, \ChARLES
% displays all.

\textbf{Steps \stepCounter{9} \& \stepCounter{10} (Visualization)} To better
understand a summary, the user requests more details \stepCounter{9}. \ChARLES
offers an interactive visualization \stepCounter{10} comprising several
non-overlapping rectangles, each representing a data partition achieved via
applying the conditions. The size of each rectangle corresponds to its data
coverage. E.g., 33.3\% employees fall within the top partition. For each
partition, additional details---such as partitioning condition, data coverage,
accuracy of the transformation---are revealed when the user hovers over these
rectangles. The bottom partition, marked by diagonal patterns, indicates that
no change was observed there.

\subsubsection*{Demonstration engagement.} Our target users are data analysts,
decision makers, and data enthusiasts who want to understand data change
trends. After our guided demonstration, participants will be able to plug their
own datasets into \ChARLES. We will also make additional
datasets~\cite{Billionaires} available. Through the demonstration, we will
showcase how \ChARLES can semantically summarize changes between two datasets.

% \af{Reduce number of references. Keep at most 10. Also, for citation 8, no need to include editor names for the venue. Also, remove any doi links from the references. You can achieve this by editing the bib file.}

\smallskip

%\clearpage

\bibliographystyle{ACM-Reference-Format}
\bibliography{paper}

\end{document}